\documentclass[aps,pra,10pt,twocolumn,showpacs,superscriptaddress]{revtex4-1}
\usepackage{hyperref}
\usepackage{amssymb}
\usepackage{amsmath}
\usepackage{epsfig}
\usepackage{graphicx}
\usepackage{subfigure}
\usepackage{dcolumn}
\newcommand{\rmm}{\mathrm{m}}
\newcommand{\rmc}{\mathrm{c}}

\newcommand{\rmL}{\mathrm{L}}
\newcommand{\rmE}{\mathrm{E}}
\newcommand{\rmY}{\mathrm{Y}}
\newcommand{\mbx}{\mathbf{x}}
\newcommand{\mbp}{\mathbf{p}}
\newcommand{\mbu}{\mathbf{u}}
\newcommand{\mbn}{\mathbf{n}}
\newcommand{\rmcm}{\mathrm{cm}}
\newcommand{\rmrp}{\mathrm{rp}}
\newcommand{\rmth}{\mathrm{th}}
\newcommand{\rmpt}{\mathrm{pt}}
\newcommand{\rmin}{\mathrm{in}}
\newcommand{\rmout}{\mathrm{out}}

\newcommand{\rmeff}{\mathrm{eff}}

\newcommand{\mcE}{\mathcal{E}}
\newcommand{\mcL}{\mathcal{L}}
\newcommand{\mcO}{\mathcal{O}}
\newcommand{\mcV}{\mathcal{V}}
\newcommand{\mcQ}{\mathcal{Q}}
\newcommand{\mcP}{\mathcal{P}}
\newcommand{\mcR}{\mathcal{R}}
\newcommand{\etal}{\textit{et al.~}}

\begin{document}
%+++++++++++++++++++++++++++++++++++++++++++++++++++++++++++++++++++++++++++++++++
\title{Quantum optomechanics of a multimode system coupled via photothermal and radiation pressure force}

\author{M.~Abdi}
%\email{m{\_}abdi@physics.sharif.ir}
\affiliation{Department of Physics, Sharif University of Technology, Tehran, Iran}

\author{A.~R.~Bahrampour}
\affiliation{Department of Physics, Sharif University of Technology, Tehran, Iran}

\author{D.~Vitali}
\affiliation{School of Science and Technology, Physics Division, University of Camerino, Camerino (MC), Italy}

\date{\today}

\begin{abstract}
We provide a full quantum description of the optomechanical system formed by a Fabry-Perot cavity with a movable micro-mechanical mirror whose center-of-mass and internal elastic modes are coupled to the driven cavity mode by both radiation pressure and photothermal force.
Adopting a quantum Langevin description, we investigate simultaneous cooling of the micromirror elastic and center-of-mass modes, and also the entanglement properties of the optomechanical multipartite system in its steady state.
\end{abstract}

\maketitle
%
%
%----------SECTION----------%
\section{Introduction}
During the last decade, cavity optomechanics has attracted the attention of a large community of physicists \cite{Naik2006,Arcizet2006,Kleckner2006,Groblacher2009}.
Micromechanical resonators mounted on such systems can be cooled down to their motional ground state~\cite{Teufel2011,Chan2011}, opening the door to the experimental study of quantum phenomena in mesoscopic mechanical objects~\cite{Favero2009,Genes2009}.
Cooling is due to an optomechanical coupling which can be realized in several ways, exploiting the radiation pressure force~\cite{Gigan2006,Arcizet2006,Kleckner2006}, the photothermal force~\cite{Metzger2004}, the optical gradient force~\cite{Lin2009}, or the Doppler force~\cite{Karrai2008}.
If the coupling becomes strong enough~\cite{Groblacher2009}, quantum mechanical correlations between the optical field and the mechanical resonator can be established, and robust stationary entanglement between optical and mechanical modes can be generated~\cite{Vitali2007,Genes2008b}.
One can exploit this entanglement for connecting various nodes of a quantum network formed by an array of similar optomechanical systems~\cite{Stannigel2010,Stannigel2012}.

In order to properly study this quantum regime, one needs a full quantum description of the optomechanical interaction. Differently from the radiation pressure
interaction, for which such a description has been developed time ago~\cite{Law1994,Law1995}, a quantum mechanical description of the photothermal force is still lacking. A semi-classical model of the photothermal force, based on a classical formalism which was developed by Metzger \etal in Ref.~\cite{Metzger2008}, was introduced and analyzed in Ref.~\cite{Pinard2008}.
Subsequent works followed a similar approach to investigate cooling~\cite{DeLiberato2011,Restrepo2011}, and more recently entanglement~\cite{Abdi2012}, in optomechanical systems where photothermal effects predominate.
Very recently, a phenomenological model for describing photothermal effects associated with the photon--phonon--exciton interaction, has been introduced by Xuereb \etal \cite{Xuereb2012}, in order to explain the experimental results by Usami~\etal that cooled a dielectric membrane by photothermal effects \cite{Usami2012}. Very few system studied the interplay of photothermal and radiation pressure effects, with the notable exception of Ref.~\cite{Marino2011} which studied the classical chaotic dynamics of the system in the presence of both effects. 

The photothermal force is generated through the following steps: i) absorption of intracavity photons by the micro-mechanical mirror which is consequently heated; ii) diffusion of the produced heat through the mirror, thus generating a temperature gradient inside the mechanical resonator; iii)
the temperature gradient acts as a source exciting thermoelastic oscillations of the internal elastic modes. However, the photothermal force acts simultaneously with the radiation pressure force, which cannot be neglected in general. This latter force couples not only the internal elastic modes, but also the mirror center-of-mass (CoM) motion, to the driven cavity mode.
Here we present a full quantum mechanical treatment of the dynamics of the multipartite optomechanical system composed by the driven cavity mode, the CoM of the mirror, and its elastic modes, based on quantum Langevin equations (QLE).
We show that under appropriate conditions, the CoM and one (or more) elastic modes of the mirror can be simultaneously cooled down close to their ground state due to the coupling provided by the photothermal and radiation pressure interaction. For stronger coupling also optomechanical entanglement at the steady state of the system can be generated, involving either the CoM mode or one elastic mode. We show that optomechanical entanglement of the CoM mode and of an elastic mode with the cavity mode cannot be simultaneously enhanced by the photothermal force, which enhances the optomechanical entanglement of the elastic mode, and tends to destroy the entanglement between the CoM mode and the cavity field.

In the Sec.~II we describe the model and derive the Hamiltonian of the system.
Dynamics of the system is investigated in Sec.~III via linearization of the QLE.
In Secs.~IV and V we exploit the steady state correlation matrix calculated in Sec.~III to examine ground state cooling and entanglement properties of the system. Sec.~VI is for concluding remarks.

%
%
%----------SECTION----------%
\section{Model}
We consider an optomechanical system composed of a Fabry--P\'erot cavity driven by an intense laser, where
one of the mirrors is a light movable mirror whose elastic normal modes and CoM oscillations around its equilibrium position can be treated as harmonic.
The micromechanical mirror absorbs the intracavity photons, and the resulting thermoelastic effect excites the internal elastic modes of the mirrors. At the same time cavity light shifts and deforms the surface of the micromirror via radiation pressure, exciting in this way both the internal elastic modes and the CoM of the mirror.

In this paper, we will study the interplay between the photothermal and the radiation pressure force, and how they affect cooling and optomechanical entanglement of both the internal and the CoM mechanical modes. In order to properly describe the system
we start from quantum elasticity theory.

\begin{figure}[b]
\includegraphics[width=\columnwidth]{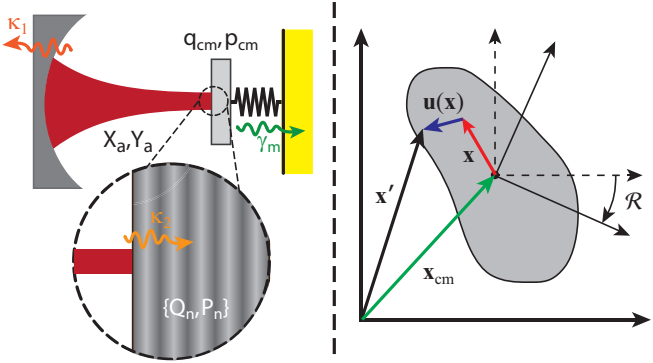}
\caption{(Color online) Scheme of the optomechanical system (left) and the geometry under consideration for an elastic system and its coordinates (right).}
\label{fig:scheme}
\end{figure}

An elastic system with density $\rho$ in the presence of an external potential $\mcV(\mbx')$ is described by a Lagrangian, given by \cite{Landau1975}
\begin{equation}\label{lagrangian}
\mcL=\int_{V}  d \mbx \Big[\frac{1}{2}\rho \dot{\mbx}'^{2} -\mcV(\mbx') -\mcV_{\rmE}(\mbx)\Big],
\end{equation}
where $\mbx'=\mbx_{\rmcm}+\mcR[\mbx+\mbu]$ is the position of an arbitrary point in the matter, in an external reference frame [see Fig.~\ref{fig:scheme}], with $\mcR$ the rotation of the object respect to the reference and $\mbu$ the displacement vector.
%%%%%%%%%%%%%%%%%%%%%%%%%%%%%%%%%%%%%%%%%%%%
%%%%%%%%%%%%%%%%%%%%%%%%%%%%%%%%%%%%%%%%%%%%
The elasticity density potential in Eq.~(\ref{lagrangian}) is given by
\begin{equation}\label{elasticity}
\mcV_{\rmE}(\mbx)=\frac{1}{2}\sum_{i,j}\varsigma_{ij}(\mbx)u_{ij}(\mbx),
\end{equation}
with elasticity tensor, $u_{ij}=\frac{1}{2}[\partial_j u_i+\partial_i u_j]$, and stress tensor
\begin{align}
\varsigma_{ij}(\mbx)=&\frac{E_{\rmY}}{1+\sigma}\big[u_{ij}(\mbx)-\frac{1}{3}\delta_{ij}\sum_{k} u_{kk}	(\mbx)\big] \nonumber \\
&+\frac{E_{\rmY}\delta_{ij}}{3(1-2\sigma)}\sum_{k} u_{kk}(\mbx),
\end{align}
where $E_{\rmY}$ and $\sigma$ are Young modulus and Poisson ratio, respectively.
By replacing $\mbx'$ into the kinetic part of the Lagrangian of Eq.~(\ref{lagrangian}), one gets \cite{Romero-Isart2011}
\begin{align}
\mcL=&\frac{1}{2}m\dot{\mbx}_{\rmcm}^{2} +\frac{1}{2}\sum_i I_i \dot{\phi}_i^2 +\frac{1}{2}\int_V 	 d\mbx \rho(\mbx)\dot{\mbu}^2(\mbx) \nonumber \\
& -\int_V  d \mbx[\mcV(\mbx')+\mcV_{\rmE}(\mbx)],
\end{align}
where $m$ is total mass of the system and the second term refers to the rotational kinetic energy.
In the present case, the system of interest is a clamped micromechanical mirror; as a consequence, the rotational degrees of freedom are not relevant and we will ignore them in the following.
The Hamiltonian of the system without any external potential reads
\begin{equation}\label{free}
H_{\rmm}=\frac{\mbp^2_{\rmcm}}{2m}+\int_V  d \mbx \big[\frac{\vec{\pi}^2}{2\rho} +\mcV_{\rmE}\big],
\end{equation}
where the conjugate momentum of the CoM and the displacement vector are defined as $\mbp_{\rmcm} \equiv \partial\mcL/\partial\dot{\mbx}_{\rmcm}$ and $\pi_i(\mbx) \equiv \partial\mcL/\partial\dot{u}_i(\mbx)$, respectively.
The equation of motion for the deformation vector, $\mbu(\mbx)$, can be easily derived from Eq.~(\ref{free}) and reads
\begin{equation}
\rho\ddot{\mbu}=\frac{E_{\rmY}\nabla(\nabla.\mbu)}{2(1+\sigma)(1-2\sigma)}+\frac{E_{\rmY}\nabla^2\mbu}{2(1+\sigma)}.
\end{equation}
One splits the vector $\mbu(\mbx)$ into the transversal and longitudinal components as $\mbu=\mbu_{l}+\mbu_{t}$, where $\nabla \cdot\mbu_{t}=0$ and $\nabla \times \mbu_{l}=0$.
Therefore, the elastic wave equations along the two directions read as follows,
\begin{align}\label{wave}
\ddot{\mbu}_l -c_l^2 \nabla^2 \mbu_l=0, \\
\ddot{\mbu}_t -c_t^2 \nabla^2 \mbu_t=0,
\end{align}
where
\begin{subequations}
\begin{eqnarray}
c_l^2 &\equiv& \frac{E_{\rmY}(1-\sigma)}{\rho(1+\sigma)(1-2\sigma)}, \\
c_t^2 &\equiv& \frac{E_{\rmY}}{2\rho(1+\sigma)},
\end{eqnarray}
\end{subequations}
are the propagation speed of the longitudinal and transversal elastic waves, respectively.

The photothermal effect can be included by noticing that when there is a temperature gradient in the material, the longitudinal wave equation attains a source term \cite{Landau1975}
\begin{equation}
\ddot{\mbu}_l -c_l^2 \nabla^2 \mbu_l =-\frac{E_{\rmY}\alpha_{\rmth}}{3\rho(1-2\sigma)}\nabla T,
\end{equation}
where $\alpha_{\rmth}$ is the linear thermal expansion coefficient of the mirror.
This source term can be taken into account by introducing an effective potential, which is ultimately caused by the absorption of the intracavity photons.
This effective thermoelastic potential adds to the effective elastic potential associated with the trapping of the CoM motion, so that, neglecting for the moment the radiation pressure effect, the total external potential is given by $\frac{1}{2}m\omega_{\rmm}^2\mbx_{\rmcm}^2 +\int_{V} d\mbx \mcV_{\rmpt}(\mbx')$,
where the photothermal potential density is given by
\begin{align}\label{photothermal}
\mcV_{\rmpt}(\mbx')&=-\frac{1}{2}\sum_{i,j}\frac{E_{\rmY}\alpha_{\rmth}\delta_{ij}}{3(1-2\sigma)}(T-T_0)u_{ij}(\mbx) \nonumber \\
&=\frac{-E_{\rmY}\alpha_{\rmth}}{6(1-2\sigma)}(T-T_0)\sum_i u_{ii}(\mbx),
\end{align}
where $T_0$ is temperature of the thermal reservoir of the mirror.
The above defined $\mcV_{\rmpt}(\mbx')$ is implicitly a function of external coordinates via the temperature term.
Actually, the temperature is a function of the number of the absorbed cavity photons, which in turn is a function of the cavity length (determined both by CoM and deformations on the surface of the mirror). The distribution of the temperature inside the micromechanical mirror can be calculated from the heat equation
\begin{equation}
\rho C \frac{\partial T}{\partial t}=K_{\rmth}\nabla^2 T+\mcQ_{\rmth},
\end{equation}
where $C$ is the specific heat capacity and $K_{\rmth}$ is the thermal conductivity of the material, while $\mcQ_{\rmth}$ is the heat flux, which in our case is due to the absorption process at the surface of the mirror.

In typical situations a one-dimensional model focusing only on longitudinal deformations of the mirror along the cavity axis $x$, and neglecting the transversal elastic modes, provide a satisfactory description of the physics. The two equations above can be rewritten in one-dimensional form as
\begin{subequations}\label{set}
\begin{eqnarray}
&&\mcV_{\rmpt}(x')=\frac{-E_{\rmY}\alpha_{\rmth}}{6(1-2\sigma)}(T-T_0) \frac{\partial u_{x}}{\partial x}, \\
&&\rho C \frac{\partial T}{\partial t}=K_{\rmth}\frac{\partial^2 T}{\partial x^2}+\frac{\beta \hbar \omega_{\rmc}}{\pi r_{0}^{2}} I_{2}^{\rmout}(x',t)\delta(x),
\end{eqnarray}
\end{subequations}
where we have explicitly written the heat flux term, with $I_{2}^{\rmout}$ the intracavity photon absorption rate, $\beta$ is quantum efficiency of the absorption, and $r_{0}$ is the radius of the optical mode on the surface of the micromechanical mirror, whose are is equal to $S$.
The Dirac delta function assures that all the photons are absorbed at the surface of the mirror \citep{Abdi2012}.
The explicit form of the effective thermolelastic potential is obtained by solving the heat equation of Eq.~(\ref{set}b) and inserting the solution into Eq.~(\ref{set}a). The heat equation must be solved within the mirror, $0<x<\ell$, where $\ell$ is the mirror thickness, with the initial condition $T(x,t=-\infty)=T_0$, and with the boundary conditions $T(x=\ell,t)=T_{0}$ and $[\partial T(x,t)/\partial x]_{x=0}=0$. The solution is~\cite{Carslaw1959}
%
%\begin{widetext}
\begin{align}\label{temperature}
T(x,t)&=T_{0}+\frac{2\beta \hbar \omega_{\rmc}}{\rho C \pi r_{0}^{2}\ell} \sum_{k=0}^{\infty}\cos\big[(2k+1)\frac{\pi x}{2\ell}\big] \\
&\times \int_{-\infty}^{t}  d t' I_{2}^{\rmout}(t') \exp\Big\{-\big[(2k+1)\frac{\pi\nu}{2\ell}\big]^{2}(t-t')\Big\}. \nonumber
\end{align}
%\end{widetext}
%
where we have defined $\nu^{2} \equiv K_{\rmth}/\rho C$.
%In fact, we have assumed that the thermoelastic effective cross-sectional area of the cantilever is the same as the optical mode.

From the homogeneous elastic wave equation of Eq.~(\ref{wave}) one can find the eigenmodes of the longitudinal elastic waves, forming a complete orthonormal set of functions, $\{u_n^0(x)\}$, normalized so that $\int_V  d \mbx ~ u_i^0(x)u_j^0(x)=V\delta_{ij}$), where $V$ is the volume of the micromechanical mirror. Therefore, an arbitrary longitudinal elastic wave, and the generalized elastic momentum as well, can be written as an expansion over this set of eigenmodes,
\begin{align}
u_x(x,t)&=\sum_n u_n^0(x)Q_n(t), \\
\pi_x(x,t)&=\frac{1}{V}\sum_n u_n^0(x)P_n(t),
\end{align}
where $P_{n}= m \dot{Q}_{n}$. At this point, it is straightforward to perform a canonical quantization by promoting the expansion coefficients $Q_{n}$ and $P_{n}$ associated with the n-th longitudinal elastic mode to operators fulfilling the canonical commutation rules $[Q_{n},P_{n}] = i\hbar$.

%++++++++++++++++++++++++++++++++++++++++++
\subsection{Hamiltonian of the system}
Using Eqs.~(\ref{free}), (\ref{set}a), and (\ref{temperature}), the total Hamiltonian of the system can be written as
$$H=H_{\rmc}+H_{\rmm}+H_{\rmrp}+H_{\rmpt},$$
where
\begin{equation}
H_{\rmc}=\hbar \omega_{\rmc} a^{\dagger}a,
\end{equation}
is the cavity mode Hamiltonian,
\begin{equation}
H_{\rmm}=\frac{1}{2}\hbar\omega_{\rmm}(p^2_{\rmcm}+q_{\rmcm}^2)
+\sum_n{\frac{1}{2}\hbar\Omega_{n}\big(P^2_n +Q_n^2\big)},
\end{equation}
is the free mechanical Hamiltonian, where we have rescaled all mechanical operators by making them dimensionless, such that $[Q_{n},P_{n}]=i$ and $[q_{\rmcm},p_{\rmcm}]=i$ and we have defined the elastic mode frequencies
\begin{equation}
\Omega_n^2 \equiv \frac{E_{\rmY}(1-\sigma)S}{\rho\ell\pi r_{0}^{2}(1+\sigma)(1-2\sigma)}\int_{0}^{\ell}  d x \big(\frac{ d u_n^0}{ d x} \big)^2.
\label{eq:freq}
\end{equation}
\begin{equation}
H_{\rmrp}=-\hbar \bigg[g_{0}q_{\rmcm} +\frac{1}{\sqrt{2}}\sum_{n}{G_{0n}Q_{n}u_{n}^{0}(x=0)}\bigg]a^{\dagger}a,
\end{equation}
is the optomechanical interaction due to radiation pressure, and
\begin{equation}
H_{\rmpt}=-\hbar G_{0n}\chi \sum_n Q_n \int_{-\infty}^{t}  d t' h_n(t-t') I_{2}^{\rmout}(t'),
\end{equation}
is the photothermal interaction term.
In the interaction terms we have introduced the radiation pressure couplings $g_{0}\equiv (\omega_{\rmc}/L)\sqrt{\hbar/m\omega_{\rmm}}$ for the CoM, $G_{0n}\equiv (\omega_{\rmc}/L)\sqrt{2\hbar/m\Omega_{n}}$ for the elastic modes, with $L$ the equilibrium cavity length, and the time-dependent photothermal couplings
\begin{subequations}\label{coupling}
\begin{eqnarray}
h_n(t) &\equiv& \frac{1}{\sqrt{2}}\int_{0}^{\ell}  d x \bigg\{(\frac{ d u_n^0}{ d x})\sum_{k}\cos\big[(2k+1)\frac{\pi x}{2\ell}\big] \nonumber \\
&&\times\exp\Big\{-\big[(2k+1)\frac{\pi\nu}{2\ell}\big]^{2}t\Big\}\bigg\}, \\
\chi &\equiv& \frac{\beta E_{\rmY}\alpha_{\rmth}L S}{3\rho C\pi r_{0}^{2} \ell(1-2\sigma)},
\end{eqnarray}
\end{subequations}
where we have singled out the dimensionless coupling coefficient $\chi$.
The derived full Hamiltonian shows that radiation pressure and the photothermal force couples the optical cavity mode simultaneously to the mirror CoM and to its internal elastic modes.
More precisely, the radiation pressure acts on both the CoM and the internal modes, while
the photothermal force excites the elastic modes only and does not affect the CoM.

In order to obtain an explicit expression for the photothermal and radiation pressure couplings and parameters, one needs the explicit form of the normalized eigenmodes $\{u_{n}^{0}(x)\}$, which is determined by the boundary conditions. In the present case both mirror surfaces are free to move, implying $du_{n}^{0}/dx=0$ at $x=0$ and $x=\ell$, so that the normalized $u_{n}^{0}$s are given by
\begin{equation}\label{un}
u_{n}^{0}(x)=\sqrt{2}\cos\big(\frac{n\pi x}{\ell}\big).
\end{equation}
Using Eq.~(\ref{coupling}a), one gets
\begin{equation}\label{response}
h_{n}(t)=\sum_{k=0}^{\infty}\dfrac{4n^{2}\exp\big\{-[(2k+1)\frac{\pi\nu}{2\ell}]^{2}t\big\}}{4n^{2}-(2k+1)^{2}},
\end{equation}
showing that in practice, the characteristic time of the photothermal force is determined by the largest thermal diffusion time, that is $\tau_{\rmth} \equiv (2\ell/\pi\nu)^{2}$.
Moreover, from Eq.~(\ref{eq:freq}) we arrive at the following explicit expression for the elastic resonance frequencies
\begin{equation}
\Omega_{n}^{2}= \frac{E_{\rmY}(1-\sigma)S}{\rho\pi r_{0}^{2}(1+\sigma)(1-2\sigma)}\Big(\frac{n\pi}{\ell}\Big)^{2}.
\end{equation}
%

%
%
%----------SECTION----------%
\section{Dynamics of the system}
Dynamics of the system can be fully characterized by its QLE, which can be derived from the Hamiltonian derived in the previous section and including damping, noise terms, and also driving of the cavity mode through the input mirror with coupling rate $\kappa_1$, by a laser with frequency $\omega_L$ and power $\mcP$. In the frame rotating at the laser frequency one has the following set of $2N+3$ of Langevin equations
\begin{subequations}\label{nonlinear}
\begin{eqnarray}
\dot{q}_{\rmcm} &=& \omega_{\rmm}p_{\rmcm}, \\
\dot{p}_{\rmcm} &=& -\omega_{\rmm}q_{\rmcm} -\gamma_{\rmm}p_{\rmcm} +g_{0}a^{\dagger}a +\xi, \\
\dot{Q}_{n} &=& \Omega_{n}P_{n}, \\
\dot{P}_{n} &=& -\Omega_{n}Q_{n} -\Gamma_{n}P_{n} +G_{0n}a^{\dagger}a \nonumber \\
&&+G_{0n}\chi\int{d t' g_{n}(t-t')I_{2}^{\rmout}(t')} +\Xi_{n}, \\
\dot{a} &=& -(\kappa_{\rmc}+i\Delta_{0})a +i\big(g_{0}q_{\rmcm}+\sum_{n}{G_{0n}Q_{n}}\big)a +\mcE \nonumber \\
&&+\sqrt{2\kappa_{1}}a_{1}^{\rmin} +\sqrt{2\kappa_{2}}a_{2}^{\rmin},
\end{eqnarray}
\end{subequations}
where in writing Eqs.~(\ref{nonlinear}d) and (\ref{nonlinear}e) we have exploited Eq.~(\ref{un}) to get $u_{n}^{0}(x=0)=\sqrt{2}$.
We have also defined $\Delta_{0}\equiv \omega_{\rmc}-\omega_{\rmL}$ and $\mcE\equiv \sqrt{2\kappa_{1}\mcP/\hbar\omega_{\rmL}}$.
The noise operators of the system are zero-mean Gaussian noises; $a_{1}^{\rmin}$ and $a_{2}^{\rmin}$ are the vacuum input noises of the cavity field entering from the two mirrors, one with decay rate $\kappa_{1}$ and the other with $\kappa_{2}$, so that the total cavity decay rate is $\kappa_{\rmc}=\kappa_{1}+\kappa_{2}$.
The correlation function of the two noises is given by
\begin{subequations}\label{input}
\begin{eqnarray}
\langle a_{i}^{\rmin}(t)a_{j}^{\rmin,\dagger}(t')\rangle &=& [\bar{N}(\omega_{\rmc})+1]\delta_{ij}\delta(t-t'), \\
\langle a_{i}^{\rmin,\dagger}(t)a_{j}^{\rmin}(t')\rangle &=& \bar{N}(\omega_{\rmc})\delta_{ij}\delta(t-t'),
\end{eqnarray}
\end{subequations}
where $i,j=1,2$ and $\bar{N}(\omega_{\rmc})\equiv \big(\exp\{\hbar\omega_{\rmc}/k_{\mathrm{B}}T\} -1\big)^{-1}$ is the mean equilibrium thermal photon number ($k_{\mathrm{B}}$ is the Boltzmann constant and $T$ is temperature of the reservoir).
At optical frequencies $\hbar\omega_{\rmc}/k_{\mathrm{B}}T \gg 1$ and therefore $\bar{N}(\omega_{\rmc})\approx 0$, so the only relevant correlation function is that of Eq.~(\ref{input}a).
The mechanical modes (CoM and the internal elastic modes) are affected by a friction force with damping rates $\gamma_{\rmm}$ for the CoM mode and $\Gamma_{n}$ for the elastic modes.
These viscous forces are associated with Brownian stochastic forces obeying the correlation function \cite{Gardiner2000}
\begin{subequations}
\begin{eqnarray}
&&\langle\xi(t)\xi(t')\rangle = \frac{\gamma_{\rmm}}{\omega_{\rmm}}\int\frac{d\omega}{2\pi}e^{-i\omega(t-t')}\omega\Big[1+\coth\big(\frac{\hbar\omega}{2k_{\mathrm{B}}T}\big)\Big], \\
&&\langle\Xi_{k}(t)\Xi_{k'}(t')\rangle = \delta_{kk'}\frac{\Gamma_{k}}{\Omega_{k}}\int\frac{d\omega}{2\pi}e^{-i\omega(t-t')}\omega \nonumber \\
&&~~~~~~~~~~~~~~~~~~~~~~~~~~~~~~~~\times\Big[1+\coth\big(\frac{\hbar\omega}{2k_{\mathrm{B}}T}\big)\Big],
\end{eqnarray}
\end{subequations}
where $k,k'=1,...,N$.

We are interested in strong optomechanical coupling, which is the optimal regime both for cooling and entanglement.
However, the single photon couplings $g_{0}$ and $G_{0n}$ are typically of the order of few KHz, and therefore they are not able to provide the required coupling strength in the weak field regime. One can circumvent this problem
by increasing the intracavity field amplitude $\langle a\rangle$, obtaining in this way an effective linearized large coupling strength.
By increasing the input power and employing high-finesse optical cavities, high intracavity intensities can be achieved.
At this situation, the cavity field is described by an intense coherent state and every mechanical mode is shifted to a new equilibrium position depending upon the number of intracavity photons.
Therefore, every operator of the system can be decomposed into two parts: a classical steady state value, and a fluctuating quantum part around this steady state value, that is, for a generic operator $\mcO$ we have $\mcO=\langle \mcO \rangle +\delta \mcO$.
By replacing in the nonlinear QLE of Eqs.~(\ref{nonlinear}), we get the steady state values as the following
\begin{align}
\langle q_{\rmcm}\rangle &= g_{0}\langle a\rangle^{2}/\omega_{\rmm},
~~~~\langle a\rangle = \frac{\mcE}{\kappa_{\rmc}+i\Delta_{\rmc}}, \nonumber \\
\langle Q_{n}\rangle &= \frac{G_{0n}\langle a\rangle^{2}}{\Omega_{n}}\Big(1 +\frac{\kappa_{2}\chi \ell^{2}}{\nu^{2}}\Big), \nonumber
\end{align}
where we have introduced the effective cavity detuning $\Delta_{\rmc}\equiv \Delta_{0} -g_{0}\langle q_{\rmcm}\rangle -\sum_{n}{G_{0n}\langle Q_{n}\rangle}$,
%We have also performed the $\int d t' h_{n}(t-t') = \ell^{2}/2\nu^{2}$, which is independent of $n$, and replaced it in (\ref{classical}d).
and we have chosen the phase of $\mcE$ can be tuned so that $\langle a\rangle$ is real and positive.

%+++++++++++++++++++++++++
\subsection{Dynamics of the quantum fluctuations}
For the small quantum fluctuations, we arrive at the following linear equations
\begin{subequations}\label{fluctuations}
\begin{eqnarray}
\delta\dot{q}_{\rmcm} &=& \omega_{\rmm}\delta p_{\rmcm}, \\
\delta\dot{p}_{\rmcm} &=& -\omega_{\rmm}\delta q_{\rmcm} -\gamma_{\rmm}\delta p_{\rmcm} +g\delta X_{a} +\xi, \\
\delta\dot{Q}_{n} &=& \Omega_{n}\delta P_{n}, \\
\delta\dot{P}_{n} &=& -\Omega_{n}\delta Q_{n} -\Gamma_{n}\delta P_{n} +\Xi_{n} +G_{n} \Big[\delta X_{a} \nonumber \\
&&+\chi \int{d t' h_{n}(t-t')\Big(2\kappa_{2}\delta X_{a} -\sqrt{2\kappa_{2}}X_{2}^{\rmin}\Big)}\Big], \\
\delta\dot{X}_{a} &=& -\kappa_{\rmc}\delta X_{a} +\Delta_{\rmc}\delta Y_{a} +\sqrt{2\kappa_{1}}X_{a1}^{\rmin} +\sqrt{2\kappa_{2}}X_{a2}^{\rmin},\\
\delta\dot{Y}_{a} &=& -\kappa_{\rmc}\delta Y_{a} -\Delta_{\rmc}\delta X_{a} +\sqrt{2\kappa_{1}}Y_{a1}^{\rmin} +\sqrt{2\kappa_{2}}Y_{a2}^{\rmin} \nonumber \\
&&+g\delta q_{\rmcm}+\sum_{n}{G_{n}\delta Q_{n}},
\end{eqnarray}
\end{subequations}
where we have introduced the cavity field quadratures $\delta X_{a}$ and $\delta Y_{a}$, so that we have $\delta a=(\delta X_{a}+i\delta Y_{a})/\sqrt{2}$ and their corresponding noise operators from each side of the cavity $X_{a1}^{\rmin}$, $X_{a2}^{\rmin}$, $Y_{a1}^{\rmin}$, and $Y_{a2}^{\rmin}$.
The effective optomechanical couplings between the quantum fluctuations are given by $g\equiv g_{0}\langle a\rangle\sqrt{2}$ and $G_{n}\equiv G_{0n}\langle a\rangle\sqrt{2}$.
Because of the integral appearing in Eq.~(\ref{fluctuations}d), we need to solve these equations in frequency space using the convolution theorem.
In the frequency domain, we can write the above equations in the following compact form
\begin{equation}\label{compact}
-\big[i\omega I +A(\omega)\big]\tilde{\mbu} = \tilde{\mbn},
\end{equation}
where $I$ is the $(2N+4)\times (2N+4)$ identity matrix and we have defined respectively the vector of system operators  $\mbu$ and the noise vector $\tilde{\mbn}$ as
\begin{align}
\mbu &\equiv \big[\delta q_{\rmcm},\delta p_{\rmcm},\delta X_{a},\delta Y_{a},\delta Q_{1},\delta P_{1},\delta Q_{2},\delta P_{2}, \cdots \big]^{\mathsf{T}}, \\
\tilde{\mbn} &\equiv \big[0,\tilde{\xi},\sqrt{2\kappa_{1}} \tilde{X}_{1a}^{\rmin} +\sqrt{2\kappa_{2}} \tilde{X}_{2a}^{\rmin},\sqrt{2\kappa_{1}} \tilde{Y}_{1a}^{\rmin} +\sqrt{2\kappa_{2}} \tilde{Y}_{2a}^{\rmin}, \nonumber \\
&~~~~~~ 0,-\sqrt{2\kappa_{2}}\chi G_{1}\tilde{h}_{1}(\omega)\tilde{X}_{2a}^{\rmin}+\tilde{\Xi}_{1},  \cdots\big]^{\mathsf{T}},
\end{align}
where $\tilde{h}_{n}(\omega)$ is Fourier transform of the response function.
By keeping only the term associated with the largest thermal diffusion time, it is straightforward to get
\begin{equation}
\tilde{h}_{n}(\omega)=\frac{4n^{2}\tau_{\rmth}}{(1 -4n^{2})(1 -i\omega\tau_{\rmth})}.
\end{equation}
The drift matrix $A(\omega)$ is given by
\begin{widetext}
\begin{equation}
A(\omega)=\left(
\begin{array}{ccccccccc}
0 & \omega_{\rmm} & 0 & 0 & 0 & 0 & 0 & 0 & \cdots \\
-\omega_{\rmm} & -\gamma_{\rmm} & g & 0 & 0 & 0 & 0 & 0 & \cdots \\
0 & 0 & -\kappa_{\rmc} & \Delta_{\rmc} & 0 & 0 & 0 & 0 & \cdots \\
g & 0 & -\Delta_{\rmc} & -\kappa_{\rmc} & G_{1} & 0 & G_{2} & 0 & \cdots \\
0 & 0 & 0 & 0 & 0 & \Omega_{1} & 0 & 0 & \cdots \\
0 & 0 & G_{1}\big(1+2\kappa_{2}\chi\tilde{h}_{1}(\omega)\big) & 0 & -\Omega_{1} & -\Gamma_{1} & 0 & 0 & \cdots \\
0 & 0 & 0 & 0 & 0 & 0 & 0 & \Omega_{2} & \cdots \\
0 & 0 & G_{2}\big(1+2\kappa_{2}\chi\tilde{h}_{2}(\omega)\big) & 0 & 0 & 0 & -\Omega_{2} & -\Gamma_{2} & \cdots \\
\vdots & \vdots & \vdots & \vdots & \vdots & \vdots & \vdots & \vdots & \ddots
\end{array}\right).
\end{equation}
\end{widetext}

The steady state associated with Eq.~(\ref{compact}) is reached when the system is stable, which occurs if and only if all the eigenvalues of $A(\omega=0)$ have negative real part.
These stability conditions can be obtained, for example, by using the Routh--Hurwitz criteria.
In this paper, we shall restrict to the situation with $\Delta_{\rmc}>0$, i.e., with a red-detuned laser, and in this parameter region the only nontrivial stability condition is
\begin{equation}\label{stability}
1-\frac{\Delta_{\rmc}}{\Delta_{\rmc}^{2}+\kappa_{\rmc}^{2}}\bigg[\frac{g^{2}}{\omega_{\rmm}}
+\sum_{n}{\frac{G_{n}^{2}}{\Omega_{n}}\Big(1+\frac{8n^{2}}{4n^{2}-1}\kappa_{2}\tau_{\rmth}\chi\Big)}\bigg]>0.
\end{equation}
%

%++++++++++++++++++++++++++++++++++
\subsection{Correlation matrix}
Since we have linearized the equations around the steady state values and all the noise operators of the system are Gaussian with zero-mean, the steady state of the quantum fluctuations is a zero-mean Gaussian state, whose properties are fully characterized by its correlation matrix.
This steady state correlation matrix is defined as
\begin{equation}
V_{ij}=\frac{1}{2}\big\langle u_{i}(\infty)u_{j}(\infty) +u_{j}(\infty)u_{i}(\infty)\big\rangle.
\end{equation}
Since at a generic time $t$ one can write
\begin{equation}
V_{ij}(t)=\iint\frac{d\omega d\omega'}{4\pi}e^{-i(\omega+\omega')t}\big\langle \tilde{u}_{i}(\omega)\tilde{u}_{j}(\omega') +\tilde{u}_{j}(\omega')\tilde{u}_{i}(\omega)\big\rangle,
\end{equation}
from the formal solution of Eq.~(\ref{compact}) one gets the following expression for the steady state correlation matrix
\begin{equation}\label{cm}
V=\int d\omega M(\omega)D(\omega)M^{\dagger}(\omega),
\end{equation}
where $M(\omega)\equiv[i\omega I +A(\omega)]^{-1}$, and we have defined the diffusion matrix $D(\omega)$ through the relation $\langle \tilde{n}_{i}(\omega)\tilde{n}_{j}(\omega') +\tilde{n}_{j}(\omega')\tilde{n}_{i}(\omega) \rangle/2 = D(\omega)\delta (\omega +\omega') $, whose explicit expression is
\begin{widetext}
\begin{equation}
D(\omega)=\left(
\begin{array}{ccccccccc}
0 & 0 & 0 & 0 & 0 & 0 & 0 & 0 & \cdots \\
0 & \dfrac{\gamma_{\rmm}\omega}{\omega_{\rmm}}\coth\big[\dfrac{\hbar\omega}{2k_{\mathrm{B}}T}\big] & 0 & 0 & 0 & 0 & 0 & 0 & \cdots \\
0 & 0 & \kappa_{\rmc} & 0 & 0 & -\kappa_{2}G_{1}\chi\tilde{h}_{1}(\omega) & 0 & -\kappa_{2}G_{2}\chi\tilde{h}_{2}(\omega) & \cdots \\
0 & 0 & 0 & \kappa_{\rmc} & 0 & 0 & 0 & 0 & \cdots \\
0 & 0 & 0 & 0 & 0 & 0 & 0 & 0 & \cdots \\
0 & 0 & -\kappa_{2}G_{1}\chi\tilde{h}_{1}(\omega) & 0 & 0 & \zeta_{1}(\omega) & 0 & \varphi_{12}(\omega) & \cdots \\
0 & 0 & 0 & 0 & 0 & 0 & 0 & 0 & \cdots \\
0 & 0 & -\kappa_{2}G_{2}\chi\tilde{h}_{2}(\omega) & 0 & 0 & \varphi_{12}(\omega) & 0 & \zeta_{2}(\omega) & \cdots \\
\vdots & \vdots & \vdots & \vdots & \vdots & \vdots & \vdots & \vdots & \ddots
\end{array}\right).
\end{equation}
\end{widetext}
and where $\zeta_{n}(\omega)$ and $\varphi_{ij}(\omega)$ are defined as
\begin{align}
\zeta_{n}(\omega)&\equiv \kappa_{2}(G_{n}\chi|\tilde{h}_{n}(\omega)|)^{2}+\frac{\Gamma_{n}}{\Omega_{n}}\omega\coth\big[\dfrac{\hbar\omega}{2k_{\mathrm{B}}T}\big], \\
\varphi_{ij}(\omega)&\equiv 2\kappa_{2}\chi^{2}G_{i}G_{j}\Re[\tilde{h}_{i}(\omega)\tilde{h}_{j}^{*}(\omega)].
\end{align}

%
%
%----------SECTION----------%
\section{Single elastic mode}
In order to see the effects of the interplay between the radiation pressure and photothermal force, we consider a situation similar to that of Ref.~\cite{Abdi2012}, i. where the detection bandwidth is chosen so that it includes the CoM mode and one elastic mode only.
In this section we focus in particular on the simultaneous cooling of the elastic mode of interest and of the CoM mode.
and also on the entanglement properties of the tripartite optomechanical system.
We choose an achievable set of system parameters, which are listed in Table~\ref{tab:table1}.
%~~~~~~~~~~~~~~~~~~~~~~~~~~~~~~~~
\begin{table}[b]
\caption{\label{tab:table1}Parameters of the optomechanical system.}
\begin{ruledtabular}
\begin{tabular}{lcc}
\textrm{Parameter}&
\textrm{Symbol}&
\textrm{Value}\\
\colrule
CoM frequency & $\omega_{\rmm}/2\pi$ & $20~\mathrm{MHz}$\\
mechanical quality factor & $Q_{\rmm}$ & $10^{5}$\\
mirror mass & $m$ & $5~\mathrm{ng}$ \\
cavity length & $L$ & $1~\mathrm{mm}$ \\
laser wavelength & $\lambda_{\rmL}$ & $810~\mathrm{nm}$ \\
reservoir temperature & $T$ & $4~\mathrm{mK}$
\end{tabular}
\end{ruledtabular}
\end{table}
%~~~~~~~~~~~~~~~~~~~~~~~~~~~~~~~~

%++++++++++++++++++++++++++++++++++++++++++
\subsection{Cooling}
Ref.~\cite{Abdi2012} has shown that photothermal effects can improve cooling of a single internal elastic mode, compared to the situation in the presence of radiation pressure only.
However, radiation pressure always couples the cavity field also to the CoM mode, and therefore for a complete description of cooling one has to include also the CoM, and to consider the simultaneous presence of the two kinds of mechanical modes.
%%%%%%%%%%%%%%%%%%%%%%%%%%%%%%%%%%%%%%%%%%%%
\begin{figure}[t]
\includegraphics[width=\columnwidth]{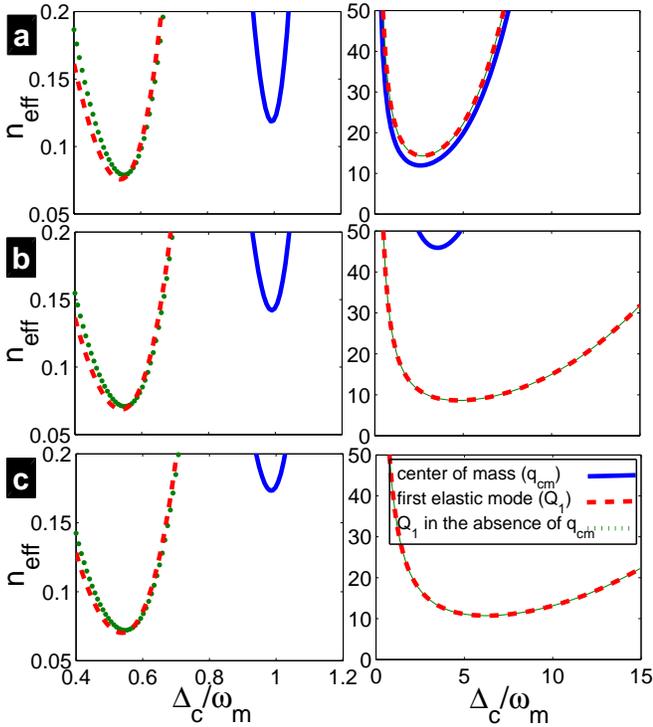}
\caption{(Color online) Effective phonon number versus cavity detuning in the good cavity limit $\kappa_{1}=0.05\omega_{\rmm}$ (left panels) and in the bad cavity limit $\kappa_{1}=5\omega_{\rmm}$ (right panels): (a) $\kappa_{2}=0$, (b) $\kappa_{2}=0.5\kappa_{1}$, and (c) $\kappa_{2}=\kappa_{1}$. The elastic mode frequency is $\Omega_{1}\simeq 0.54\omega_{\rmm}$, $\tau_{\rmth}=\Omega_{1}^{-1}$ ,and $\chi \simeq 0.13$. Power of the laser is $\mcP=15$~mW.}
\label{fig:cooling11}
\end{figure}
%%%%%%%%%%%%%%%%%%%%%%%%%%%%%%%%%%%%%%%%%%%%

In Fig.~\ref{fig:cooling11} the effective phonon number of the CoM mode (blue full line) and of the elastic modes of interest (red dashed line) is plotted versus the effective cavity detuning $\Delta_{\rmc}$, for two different input rates $\kappa_{1}$, corresponding to two completely different regimes: the resolved side-band regime ($\kappa_{1}=0.05\omega_{\rmm}$, left panel) and the bad cavity regime ($\kappa_{1}=5\omega_{\rmm}$, right panel).
We consider three photothermal strengths:(a) $\kappa_{2}=0$ (only radiation pressure), (b) $\kappa_{2}=0.5\kappa_{1}$, and (c) $\kappa_{2}=\kappa_{1}$.
From the figure, one can easily conclude that including photothermal effects considerably influences the cooling of the mechanical modes of the system both in the good and in the bad cavity regimes.
In fact, while slightly improving cooling of the elastic mode, photothermal force worsens the ground state cooling of the CoM.
This may be explained in terms of the increase of the cavity mode bandwidth caused by photon absorption in the micromirror.

This effect is significant in the bad cavity limit, so that we cannot even see the curve of the CoM phonon number in right panel of Fig.~\ref{fig:cooling11}(c).
In Fig.~\ref{fig:cooling11}, we have also plotted $n_{\rmeff}$ of the elastic mode in the absence of CoM mode (same as the case of Ref.~\cite{Abdi2012}).
In the case of $\kappa_{\rmc}\gg \omega_{\rmm}$ the absence or presence of the CoM mode does not change the situation, so the two curves coincide.
However, in the resolved sideband limit, one can see that the presence of the CoM mode slightly improves cooling of the elastic mode toward its ground state.

\begin{figure}[b]
\includegraphics[width=\columnwidth]{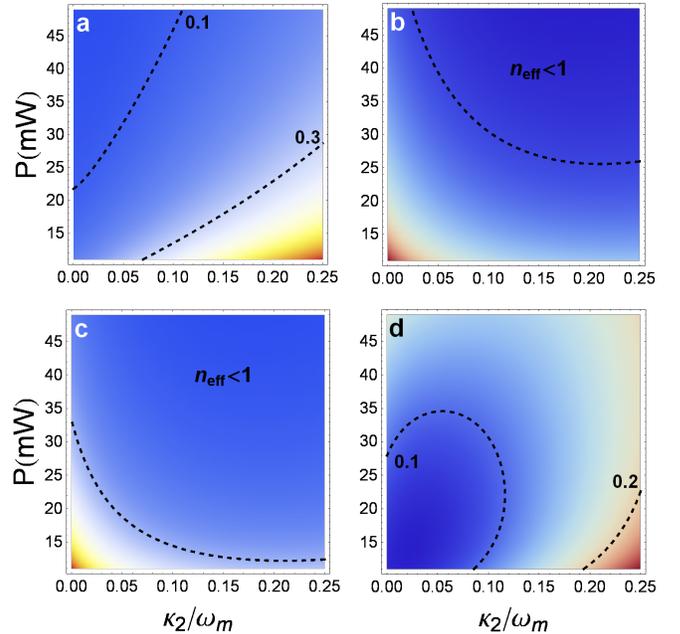}
\caption{(Color online) Density plots of $n_{\rmeff}$ as a function of $\kappa_{2}$ and laser power. (a) and (c) refer to the CoM mode, (b) and (d) to the elastic mode. The upper panels refer to $\Delta_{\rmc}=\omega_{\rmm}$, while the lower panels are for $\Delta_{\rmc}=\Omega_{1}$. The cavity input rate is $\kappa_{1}=0.05\omega_{\rmm}$. The other parameters are the same as in Fig.~\ref{fig:cooling11}.}
\label{fig:cooling12}
\end{figure}

In Fig.~\ref{fig:cooling12}, we present the density plot of the effective phonon number of the CoM mode and of the elastic mode versus laser input power and absorption rate when the cavity detuning is set to $\Delta_{\rmc}=\omega_{\rmm}$ (top panels) and $\Delta_{\rmc}=\Omega_{1}$ (low panels).
For the CoM mode, increasing the laser power improves the cooling process; at fixed power the way photothermal force affects cooling of the CoM mode depends upon the cavity detuning: for a cavity tuned to $\omega_{\rmm}$ the bolometric force has a detrimental effect on cooling (Fig.~\ref{fig:cooling12}(a)), 
while, for a cavity tuned to $\Omega_{1}$, that is to the motional sideband associated with the internal mode, cooling improves for increasing absorption losses. However the relevant parameter for simultaneous cooling is just the total cavity decay rate $\kappa_c$. In fact,
the interval for the detuning $\Delta_{\rmc}$ within which one has a significantly low value of $n_{\rmeff}$ for the CoM mode is given by $\omega_{\rmm}-\kappa_{\rmc} \lesssim \Delta_{\rmc} \lesssim \omega_{\rmm}+\kappa_{\rmc}$, consistent with the expression of the net laser cooling rate in the presence of radiation pressure interaction~\cite{Wilson-Rae2007,Marquardt2007,Genes2008}.
By increasing the absorption rate, the total cavity decay rate $\kappa_c$ increases, and the mechanical modes falling within this bandwidth are cooled, even for small laser powers (see also Ref.~\cite{Genes2008a}).
%%%%%%%%%%%%%%%%%%%%%%%%%%%%%%%%%%%%%%%%%%%%
%%%%%%%%%%%%%%%%%%%%%%%%%%%%%%%%%%%%%%%%%%%%

\begin{figure}
\includegraphics[width=\columnwidth]{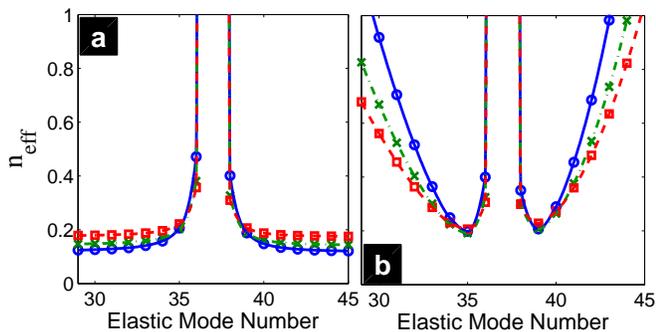}
\caption{(Color online) Effective phonon number of the CoM mode (a) and of the elastic mode (b), as a function of the elastic mode frequency for three different photon absorption rate values: $\kappa_{2}=0$ (blue circle), $\kappa_{2}=0.5\kappa_{1}$ (green cross), and $\kappa_{2}=\kappa_{1}$ (red square). The detuning is set to $\Delta_{\rmc}=\omega_{\rmm}$ and $\chi\simeq 0.03$. Laser power is $\mcP=15$~mW and $\kappa_{1}=0.05\omega_{\rmm}$.}
\label{fig:cooling13}
\end{figure}

In Fig.~\ref{fig:cooling13} we show $n_{\rmeff}$ for the CoM mode and for a selected internal elastic mode as a function of the internal mode index $n$, and therefore of the elastic resonance frequencies $\Omega_{1}(n)$. Blue circles refer to the case when the optomechanical interaction is provided only by radiation pressure, while green crosses and red squares refer to two different photothermal strengths, $\kappa_{2}=0.5\kappa_{1}$ and $\kappa_{2}=\kappa_{1}$, respectively.
On the one hand, cooling for the CoM mode, is achieved both with and without the photothermal effect, provided that the elastic mode frequency is not to close to the CoM frequency $\omega_{\rmm}$.
On the other hand, cooling of the elastic mode improves for increasing absorption losses $\kappa_{2}\neq 0$, even though is optimal only
for elastic resonance frequencies not too far (and not too close) from the CoM frequency $\omega_{\rmm}$. Within there two narrow intervals around $\omega_{\rmm}$ in fact, the elastic mode frequency satisfies the optimal resonance condition $\Delta_{\rmc}\sim \Omega_{1}(n)$, but it is not too close to $\omega_{\rmm}$ where a destructive interference phenomenon between the two cooling processes takes place preventing simultaneous cooling, as explained in detail in~\cite{Genes2008a}. 
In the present case this occurs when $\Omega_{1}(n=37)\simeq \omega_{\rmm}$, where cooling is practically absent.

%%%%%%%%%%%%%%%%%%%%%%%%%%%%%%%%%%%%%%%%%%%%
%%%%%%%%%%%%%%%%%%%%%%%%%%%%%%%%%%%%%%%%%%%%

%++++++++++++++++++++++++++++++++++++++++++
\subsection{Entanglement}
We now study the entanglement properties of the tripartite system formed by the CoM mode, a selected elastic mode, and the cavity mode, by focusing
on the entanglement of the three possible bipartite systems at the steady state.
As a measure of bipartite entanglement we choose the logarithmic negativity~\cite{Simon2000,Vidal2002,Adesso2004}, given by $E_{N}=\mathrm{max}\big[0,-\ln 2\eta^{-}\big]$, where $\eta^{-}$ is the minimum symplectic eigenvalue of the partially transposed bipartite covariance matrix \cite{Vidal2002}.

\begin{figure}[b]
\includegraphics[width=\columnwidth]{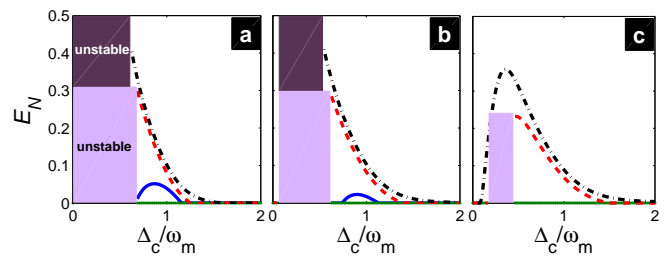}
\caption{(Color online) Logarithmic negativity of the three distinct bipartite subsystems: CoM mode--cavity mode (blue solid line), elastic mode--cavity mode (red dashed line), CoM mode--elastic mode (green dotted line), and elastic mode--cavity mode when the CoM mode is absent (black dot-dashed line), as a function of $\Delta_{\rmc}$; (a) $\kappa_{2}=0$, (b) $\kappa_{2}=0.5\kappa_{1}$, and (c) $\kappa_{2}=\kappa_{1}$. Laser power is $\mcP=50$~mW, $\kappa_{1}=0.25\omega_{\rmm}$, $\Omega_{1}\simeq 0.54\omega_{\rmm}$, $\chi \simeq 0.13$, and $\tau_{\rmth}=1/\Omega_{1}$.}
\label{fig:entang11}
\end{figure}

We first investigate the dependence of the logarithmic negativities upon the effective cavity detuning at different values of the cavity absorption losses $\kappa_{2}$ (see Fig.~\ref{fig:entang11}).
In Fig.~\ref{fig:entang11} we have also included for comparison the negativity of the elastic mode--cavity mode bipartite system in the absence of the CoM mode (black dot-dashed line).
Since the stability condition in this latter case (\textit{i}) differs from the stability conditions of the tripartite system including also the CoM mode (\textit{ii}), we have used different colors ---dark purple for case (\textit{i}) and light purple for (\textit{ii})--- to denote the unstable region in each case.
One can see from Fig.~\ref{fig:entang11} that when there is no photon absorption ($\kappa_{2}=0$), the presence of CoM mode does not appreciably affect elastic mode--cavity mode entanglement, and only the instability of the system prevents this bipartite optomechanical entanglement to achieve values large than those achieved without the CoM (\textit{i}).
In the presence of a nonzero photon absorption rate, decoherence of both systems is increased, and therefore the bipartite $E_{N}$'s decrease (see Figs.~5(b) and (c)).
However, while for the case without the CoM (\textit{i}) the stable region rapidly extends, allowing the $E_{N}$ of the elastic mode--cavity mode subsystem to reach higher values,  in the presence of the CoM (\textit{ii}) the stability region is not significantly altered by the photothermal force, and the $E_{N}$ cannot reach similar values (see Fig.~\ref{fig:entang11}(b)). At the same time a small but nonzero  CoM--cavity mode entanglement in a narrow interval of detuning is found.
Further increase of photon absorption $\kappa_{2}$ has a detrimental effect of the bipartite optomechanical entanglement, so that there remains no CoM--cavity mode entanglement when $\kappa_{2}=\kappa_{1}$, while the elastic mode--cavity mode entanglement decreases even though it remains non-negligible.
We notice in addition that the CoM mode--cavity mode $E_{N}$, differently from the elastic mode---cavity mode $E_{N}$, is not maximum at the bistability threshold, and that by increasing the photothermal force the first region at which the entanglement disappears is just at the edge of instability region.

We finally notice that for the parameter region considered here, there is no entanglement between the two mechanical modes. Here we found results consistent with those of Ref.~\cite{Genes2008a}, which showed that small but nonzero bipartite mechanical entanglement can be obtained only in a regime where the
oscillators are heavily damped and the cavity finesse is very high ($\kappa_{\rmc}\sim \gamma_{\rmm}<\omega_{\rmm}$).
%%%%%%%%%%%%%%%%%%%%%%%%%%%%%%%%%%%%%%%%%%%%
%%%%%%%%%%%%%%%%%%%%%%%%%%%%%%%%%%%%%%%%%%%%

\begin{figure}
\includegraphics[width=\columnwidth]{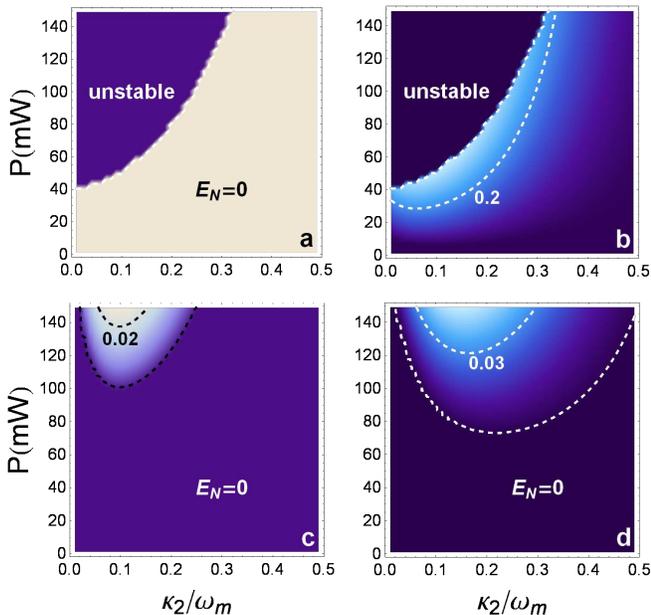}
\caption{(Color online) Density plot for the logarithmic negativity of CoM mode--cavity mode (a) and (c) , and of elastic mode--cavity mode bipartite subsystems (b) and (d), as a function of $\kappa_{2}$ and laser power. Cavity detuning is $\Delta_{\rmc}=0.4\omega_{\rmm}$ for the upper panels and $\Delta_{\rmc}=0.9\omega_{\rmm}$ for the lower panels, and the cavity input rate is $\kappa_{1}=0.05\omega_{\rmm}$. The other parameters are the same as in Fig.~\ref{fig:entang11}.}
\label{fig:entang12}
\end{figure}

To see that to what extent the photothermal force influences bipartite optomechanical entanglement and to examine its relation with the stability conditions, we plot $E_{N}$ of the CoM mode--cavity mode and elastic mode--cavity mode bipartite subsystems as a function of the absorption rate $\kappa_{2}$ and of the power of the pump laser ($E_{N}$ of the bipartite subsystem composed of the mechanical modes is again zero in the whole parameter region).
From Fig.~\ref{fig:entang12}(a) and (b), which refers to $\Delta_{\rmc}=0.4\omega_{\rmm}$, one can easily see that the presence of photothermal effects leads to a narrowing of the instability region.
As a consequence, while the maximum value of elastic mode--cavity mode $E_{N}$ is about $0.21$, when $\kappa_{2}=0$, by increasing the photon absorption rate, the instability threshold is ``pushed back'' and the elastic mode--cavity mode$E_{N}$ may reach values as high as $0.26$.
On the contrary, the CoM mode--cavity mode $E_{N}$ at this value of the detuning $\Delta_{\rmc}$ is always absent.
Instead, for $\Delta_{\rmc}=0.9\omega_{\rmm}$ which refers to Fig.~\ref{fig:entang12}(c) and (d), one can achieve a small value of the optomechanical entanglement for both the CoM and the elastic mode, at high pump powers.
We notice from Fig.~\ref{fig:entang12}(c) and (d) that there is an optimal nonzero value of $\kappa_{2}$ for both the CoM mode--cavity mode and elastic mode--cavity mode entanglement giving a maximum value for $E_{N}$. This fact occurs also for the CoM mode even though it is not affected by photothermal effects, because the entanglement depend upon the total cavity decay rate $\kappa_{\rmc}=\kappa_{1}+\kappa_{2}$ which must be not too small in order to have an appreciable entanglement.
%%%%%%%%%%%%%%%%%%%%%%%%%%%%%%%%%%%%%%%%%%%%
%%%%%%%%%%%%%%%%%%%%%%%%%%%%%%%%%%%%%%%%%%%%

\begin{figure}
\includegraphics[width=\columnwidth]{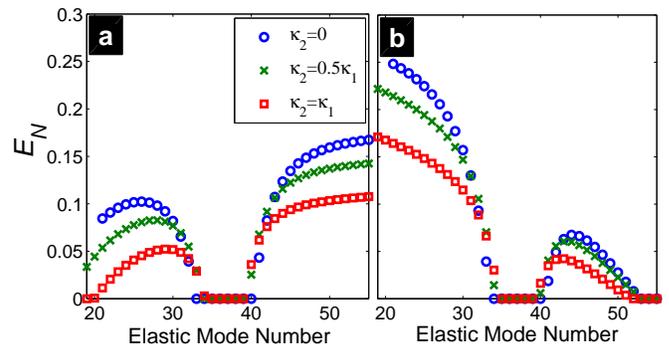}
\caption{(Color online) Logarithmic negativity of CoM mode--cavity mode (a), and elastic mode--cavity mode (b) bipartite subsystems as a function of the elastic mode frequency. The detuning is set to $\Delta_{\rmc}=0.87\omega_{\rmm}$ and $\chi\simeq 0.03$. Laser power is $\mcP=100$~mW and $\kappa_{1}=0.25\omega_{\rmm}$.}
\label{fig:entang13}
\end{figure}

Similarly to what we have done for cooling, we finally analyze the behavior of bipartite optomechanical entanglement as a function of the difference between the CoM and elastic mechanical frequencies.
In Fig.~\ref{fig:entang13} we plot CoM mode--cavity mode (Fig.~\ref{fig:entang13}(a)) and elastic mode--cavity mode (Fig.~\ref{fig:entang13}(b)) entanglement versus the elastic mode index number which determines the value of $\Omega_{1}$, for three different values of the intracavity photon absorption rate $\kappa_{2}$.
Similarly to the results of Ref.~\cite{Genes2008a}, when the frequency of the elastic mode is close enough to the CoM frequency ($\Omega_{1}(n=37)\approx \omega_{\rmm}$), the two optomechanical entanglements vanish. In fact, as illustrated in Ref.~\cite{Genes2008a},
when the two mechanical modes are at resonance, the cavity mode is strongly coupled, and entangled, to the collective coordinate $(g_{0}q_{\rmcm}+G_{01}Q_{1})/(g_{0}^{2}+G_{01}^{2})$, and uncoupled from the relative coordinate $(g_{0}q_{\rmcm}-G_{01}Q_{1})/(g_{0}^{2}+G_{01}^{2})$.
The CoM mode and the single elastic mode are linear combinations of these two coordinates and therefore their entanglement depends upon the state of these coordinates.
At $T = 0$, the CoM mode and the elastic mode of interest are entangled with the cavity mode, due to the strong coupling of the collective coordinate with the cavity mode and because the relative coordinate, even though uncoupled, is in its ground state and does not degrade the established optomechanical entanglement.
However, as soon as temperature is increased, the thermal fluctuations of the uncoupled relative coordinate prevails over the effect of the optomechanical coupling and destroys the entanglement of the CoM and of the elastic mode with the cavity field.
%%%%%%%%%%%%%%%%%%%%%%%%%%%%%%%%%%%%%%%%%%%%
%%%%%%%%%%%%%%%%%%%%%%%%%%%%%%%%%%%%%%%%%%%%

Furthermore, it is interesting to note that for the elastic modes at optimal distance from the CoM mode ($31\leq n \leq 34$ and $40\leq n \leq 43$) a the presence of photothermal coupling gives higher optomechanical entanglement with respect to the situation with only the radiation pressure interaction.
Fig.~\ref{fig:entang13} also show that entanglement sharing also occurs in the steady state of the tripartite system under study.
In fact, the elastic mode---cavity mode entanglement is redistributed to the CoM mode---cavity mode entanglement as soon as radiation pressure coupling and therefore the CoM is present. More in detail, when $\Omega_{1}<\omega_{\rmm}$ the elastic mode--cavity mode $E_{N}$ is much greater than CoM mode--cavity mode entanglement, while for $\Omega_{1}>\omega_{\rmm}$ the opposite situation occurs.
Hence, one can increase the optomechanical entanglement of one mode at the expense of the other by selecting the appropriate internal elastic mode.
%
%
%----------SECTION----------%
\section{Conclusion}
In conclusion, we have introduced a general Hamiltonian including both photothermal and radiation pressure interactions in an optomechanical system formed by a Fabry-Perot cavity with a movable micromirror. Even if referred to this specific system, one could apply this analysis to different optomechanical devices and geometries. 
The two kind of interactions couple a driven cavity mode to the CoM mode of the micromirror and to several internal elastic modes, and we have analyzed the quantum properties of this optomechanical system, by adopting a QLE approach.
By linearizing the QLE of the system, cooling of the mechanical modes has been studied:
we have shown that in addition to the CoM mode, one can cool the internal elastic modes of the mirror toward their ground state. This simultaneous cooling is helped by the photothermal force in the sense that the parameter region where simultaneous cooling is achieved is extended by its presence.
As first shown in Ref.~\cite{Genes2008a}, cooling vanishes because of a classical destructive interference only when the CoM and the elastic mode mechanical frequencies becomes close to each other.
We have then studied bipartite entanglement of the tripartite system formed by the CoM mode, a selected internal elastic mode and the cavity mode, focusing especially on the two optomechanical bipartite systems, because entanglement between the two mechanical modes is achievable only in a very narrow set of parameter space.
As expected due to entanglement monogamy, the optomechanical entanglement of CoM mode and elastic modes with the cavity mode cannot be simultaneously enhanced by the photothermal force.
In fact, the bolometric force enhances the optomechanical entanglement of the elastic mode, while destroying the entanglement between the CoM mode and the cavity mode.

%---------------------------------------------------------------------------------
\bibliography{elasticity}
%+++++++++++++++++++++++++++++++++++++++++++++++++++++++++++++++++++++++++++++++++
\end{document}